\RequirePackage{fix-cm}
\documentclass[smallextended]{svjour3}       
\smartqed  
\usepackage{times}
\usepackage{graphicx}
\usepackage{epsf}
\usepackage[colorlinks={true}]{hyperref}
\usepackage{amssymb}
\usepackage[T1]{fontenc}
\usepackage[utf8]{inputenc}
\hypersetup{citecolor={blue}, filecolor={blue}, linkcolor={blue}, urlcolor={blue}}

\newcommand{\commentold}[1]{}
\DeclareMathSymbol{:}{\mathpunct}{operators}{"3A}
\begin{document}

\title{Quantum speed limit time for correlated quantum channel
}


\author{N. Awasthi$^{1}$,  S.  Haseli$^{2}$, U. C. Johri$^{3}$, S. Salimi$^{4}$ , H. Dolatkhah$^{4}$ , A. S. Khorashad$^{4}$ 
}


\institute{N. Awasthi$^{1}$ \at $^{1}$ Department of Physics, DIT University Mussoorie, Diversion Road, Makka Wala, Uttarakhand 248009, India.\\
S.  Haseli$^{2}$ \at
              $^{2}$Faculty of Physics, Urmia University of Technology, Urmia, Iran. \\
              \email{soroush.haseli@uut.ac.ir}    \\       
              U. C.  Johri$^{3}$ \at $^{3}$College of Basic Sciences and Humanities, G.B. Pant University Of Agriculture and Technology, Pantnagar, Uttarakhand - 263153, India. \\
           \and
           S. Salimi$^{4}$ , H. Dolatkhah$^{4}$ , A. S. Khorashad$^{4}$ \at
              $^4$Department of Physics, University of Kurdistan, P.O.Box 66177-15175, Sanandaj, Iran.
}

\date{Received: date / Accepted: date}

\maketitle

\begin{abstract}
Memory effects play a fundamental role in the dynamics of open quantum systems. \textbf{There exist two different views on memory for quantum noises}. \textbf{In the first view, the quantum channel has memory when there exist correlations between successive uses of the channels on a sequence of quantum systems}. These types of channels are also known as correlated quantum  channels. \textbf{In the second view, memory effects result from correlations which are created  during the quantum evolution}.  \textbf{In this work we will consider the first view and study the quantum speed limit time for a correlated quantum channel. Quantum speed limit time is the bound on the minimal time which is needed for a quantum system to evolve from an initial state to desired states. The quantum evolution is fast if the quantum speed limit time is short.} In this work, we will study the quantum speed limit time for some correlated unital and correlated non-unital channels. As an example for unital channels we choose correlated dephasing colored noise. We also consider the correlated amplitude damping and correlated squeezed generalized amplitude damping channels as the examples for non-unital channels. \textbf{It will be shown 
 that the quantum speed limit time for correlated  pure dephasing colored noise is increased by increasing correlation strength, while for correlated amplitude damping and correlated squeezed generalized amplitude damping  channels quantum speed limit time is decreased by  increasing correlation strength. } 
\end{abstract}

\section{Introduction}	
It is impossible to isolate a quantum system from its surroundings leading to information loss in the form of dissipation and decoherence. The theory of open quantum systems offers the necessary tools for describing and analyzing the interactions of a system with its  surrounding \cite{Breuer}. In the theory of open quantum systems, various methods have been proposed to illustrate the environment and its effects on the dynamics of the desired system \cite{Breuer,Weiss}.  If the coupling strength between the system and the environment is weak and the relaxation time of the system is longer than  correlation time of the environment,  then there exist a one-way flow of information from the system to the environment. Such a quantum evolution is called Markovian \cite{Gorini} and it can be described by the  master equation in  Lindblad form \cite{Lindblad,Zhang1,Carmichael}. In a more realistic situation, the coupling strength between the system and the environment is strong and  the relaxation time of the  system is shorter than the correlation time of the environment. In this case there exist a back-flow of information from environment to the system. This type of quantum evolution is called non-Markovian\cite{Wolf,Rivas,Rivas1,Haseli,Haseli1,Haseli2,Fanchini}. In the theory of open quantum systems, dynamical memory effects play a fundamental role in various physical phenomena such as quantum biology \cite{Lambert,Thorwart,Huelga}, quantum cryptography \cite{Vasile}, quantum metrology \cite{Chin} and quantum control \cite{Hwang}. \textbf{ According to the types of dynamical memory effects, quantum evolutions can be divided  into two categories:  memory-less evolution (Markovian evolution) and  quantum process with memory (non-Markovian evolution)}. \textbf{In the case of non-Markovian evolution, future states of a system can be depend on its past because of the  back-flow of information . So it is natural to conclude that the back-flow of information from the environment to the system has a direct relation to the existence of memory.} 

This view about the dynamical memory effects and non-Markovianity as a typical part of the theory of open quantum systems is completely different from the concept of quantum channels with memory. In order to distinguish between these two, the term "correlated quantum channel" is used to describe quantum channels with memory. The memory of the quantum channel is depicted by successive uses of the channels on a sequence of quantum systems \cite{Macchiavello,Caruso,Kretschmann}. \textbf{In this sense, channels with memory and without memory channels represent cases in which the successive uses of  the channels are correlated or independent, respectively}. In the case of a correlated quantum channel, memory is not due to the correlations created during the time evolution but due to the correlated action of channels on the system  consisting of a set of individual quantum systems. Addis et al. have studied  the connection  between these two insight about memory   in Ref. \cite{Addis}.

The dynamics of quantum correlations under correlated quantum channels has been studied previously. In Ref. \cite{Ramzan}, the effects of correlated quantum channels have been investigated on the entanglement of $X$-type state of the Dirac fields in the non-inertial frame. In Ref. \cite{Guo}, the authors have shown that how the  correlated channel affects the dynamics of quantum correlations. The behavior of memory-assisted  entropic uncertainty relation under the effects of the correlated   quantum  channels has been investigated in Refs. \cite{Karpat,Guo1}. 

We study the quantum speed limit QSL time for correlated and uncorrelated quantum channels. QSL time  is the bound on the minimal time which is needed for a quantum system to evolve from an initial state at initial time $\tau$ to desired states at time $\tau+\tau_D$, where $\tau_D$ is the driving time. In Ref.\cite{Mandelstam}, Mandelstam and  Tamm have provided a bound for closed quantum systems which is given by 
\begin{equation}\label{MT1}
\tau \geq \tau_{QSL}=\frac{\pi \hbar}{2 \Delta E},
\end{equation}
where $\Delta E = \sqrt{\langle \hat{H}^{2} \rangle - \langle \hat{H} \rangle^{2}}$ is the inverse of the variation of energy of the initial state and $\hat{H}$ is time-independent Hamiltonian  describing the dynamics of quantum system. This bound is known as the MT bound. Margolus and  Levitin have presented  the bound for closed quantum system based on the mean energy $E=\langle \hat{H} \rangle$ as \cite{Margolus} 
\begin{equation}\label{ML1}
\tau \geq \tau_{QSL}=\frac{\pi \hbar}{2  E},
\end{equation}
which is called the ML bound. Combining the MT and ML bounds in Eqs. (\ref{MT1}) and (\ref{ML1}) provides a unified bound for the QSL time for the dynamics of closed  quantum system as \cite{Giovannetti}
\begin{equation}
\tau \geq \tau_{QSL}=\max \lbrace  \frac{\pi \hbar}{2 \Delta E} , \frac{\pi \hbar}{2  E} \rbrace.
\end{equation}
Recently, QSL time has also been studied for the dynamics of open quantum systems which are described by positive non-unitary maps.  For open quantum systems QSL time has quantified based on quantum Fisher information  \cite{Taddei,Escher}, Bures angle \cite{Deffner1}, relative purity \cite{del,Zhang} and other proper geometric approach  \cite{Xu,Mondal,Levitin,Xu1,Meng,Mirkin,Campaioli,Uzdin}.

In this work, we will show how correlations in the application of quantum channels can effect QSL time.  We provide  results for  some  unital and non-unital correlated channels. \textbf{We will consider   random dephasing correlated noise as an example for unital correlated quantum channels and consider correlated amplitude  damping and correlated squeezed generalized amplitude damping channels SGAD as  examples for the non-unital correlated quantum channels. }

\textbf{In this work,  we investigate the effects of correlations in the quantum channel on QSL time, so we do not limit ourselves to choose a particular measure of QSL time.  We use the bound based on relative purity for QSL time which was introduced in \cite{Zhang}}. \textbf{We choose this bound because it can be used for arbitrary initial pure and mixed states}.  

This work is organized as follows. In Sec. \ref{Sec.2}  we review the geometric approaches based on relative purity  for driving the QSL bounds. In Sec. \ref{Sec3}, the QSL time for correlated and uncorrelated quantum channel is investigaed. We will consider correlated pure dephasing colored noise as an example for correlated unital channel and consider correlated amplitude damping and squeezed generalized amplitude damping SGAD channels as examples for the non-unital correlated quantum channels. In Sec. \ref{Sec4}, we summarize the results. 


\section{The quantum speed limit time for open quantum system}\label{Sec.2}
\textbf{The state of the open quantum system at time $t$ is characterized by density matrix $\rho_t$. Time evolution of  an open quantum system is defined by the following time-dependent master equation as
\begin{equation}
\dot{\rho}_{t}=L_{t}(\rho_{t}),
\end{equation}
where $L_{t}$ is the positive generator  \cite{Breuer}. The goal is to find  the minimum time for evolving from the state $ \rho_{\tau}$ at time $\tau$  to desire state $\rho_{\tau + \tau_D}$ at time $\tau + \tau_D$. Here, $\tau_D$ is the driving time of the open quantum system. Based on relative purity the QSL time has been introduced by the authors in Refs. \cite{del,Zhang}. Zhang et al. have shown that this  QSL time is applicable for arbitrary initial mixed and pure states.  The relative purity $f(\tau)$ between initial state $\rho_{\tau}$ and desire state $\rho_{\tau+\tau_D}$ is given by \cite{Audenaert}
\begin{equation}\label{relative purity}
f(\tau + \tau_D)=\frac{tr(\rho_{\tau}\rho_{\tau + \tau_D})}{tr(\rho_{\tau}^{2})}.
\end{equation}
 The ML bound of QSL time  for open quantum systems is given by (see Ref. \cite{Zhang} for details)
\begin{equation}\label{ML}
\tau \geq \frac{\vert f( \tau + \tau_D ) -1 \vert tr (\rho_{\tau}^{2})}{\overline{ \sum_{i=1}^{n} \Lambda_{i} \beta_{i}}},
\end{equation}
where $\Lambda_{i}$ and $\beta_{i}$ are the singular values of $\mathcal{L}_{t}(\rho_{t})$ and $\rho_{\tau}$, respectively and in the denominator of the bound, we have  $\overline{\square}=\frac{1}{\tau_{D}} \int_{\tau}^{\tau + \tau_{D}} \square dt$. Following the same procedure the MT bound of QSL-time for open quantum systems  can be written as 
\begin{equation}\label{MT}
\tau \geq \frac{\vert f( \tau + \tau_D ) -1 \vert tr (\rho_{\tau}^{2})}{\overline{ \sqrt{\sum_{i=1}^{n} \Lambda_{i}^{2}}}}.
\end{equation}
Combining Eqs. (\ref{ML}) and (\ref{MT}) leads to a unified bound for QSL time as
\begin{equation}\label{(QSL)T}
\tau_{QSL}=\max \lbrace \frac{1}{\overline{ \sum_{i=1}^{n} \Lambda_{i} \beta_{i}}}, \frac{1}{\overline{ \sqrt{\sum_{i=1}^{n} \Lambda_{i}^{2}}}} \rbrace \times \vert f( \tau + \tau_D ) -1 \vert tr (\rho_{\tau}^{2}).
\end{equation}
Zhang et al. have  shown that the QSL-time is associated  with quantum coherence of an arbitrary initial state $\rho_{\tau}$ \cite{Zhang}. They have also shown that for open quantum systems the ML  bound of the QSL time  in Eq. (\ref{ML}) is tighter than MT bound. QSL time is shorter than $\tau_D$. It is worth noting that QSL-time $\tau_{(QSL)}$ can be interpreted as the potential capacity for further evolution acceleration. If $\tau_{QSL}=\tau_{D}$ then the evolution is now in the situation with the highest speed, thus the evolution does not have the potential capacity for further acceleration. However, when $\tau_{QSL} < \tau_D$, the potential capacity for further acceleration will be greater. Another important point to be noted is that when the coupling strength  between the system and environment is weak $\tau_{QSL}$ tends to the actual driving time $\tau_{D}$. On the contrary, in the  strong coupling  limit between the system and environment, $\tau_{QSL}$ can  reduce below the actual driving time $\tau_D$ \cite{Deffner1}.}

\section{Correlated quantum channels}\label{Sec3}
We provide a brief review on quantum channel with correlated noise \cite{Macchiavello,Caruso,Kretschmann,Addis,Ramzan,Guo,Yeo,Awasthi}. Quantum channels are divided into two categories of with memory and without memory channels. If the correlation time of the environment is shorter than the time between successive application then there is no correlation between consecutive uses of  the channel, i.e. a quantum channel $\varepsilon$  for $N$ consecutive uses obey $\varepsilon_{N}=\varepsilon^{\otimes N}$. These kinds of channels are known as channel without memory (uncorrelated channels). However, in real physical quantum noise, it is logical to have correlations between consecutive application  of the channels. In this case, the correlation
time of the environment is longer than the time between the successive application of the channels, i.e. a quantum channel $\varepsilon$  for $N$ successive uses  obeys $\varepsilon_{N} \neq \varepsilon^{\otimes N}$. These kinds of channels are known as memory channels (correlated channels). For correlated channels, the channel acts dependently on each input. Here we consider $N$  consecutive uses of the quantum channel. A quantum channel $\varepsilon$ can be represented as a completely positive, trace-preserving map from input state $\rho$ to output $\varepsilon(\rho)$ in Kraus form
\begin{equation}
\varepsilon(\rho)=\sum_{i_{1}...i_{N}}E_{i_{1}...i_{N}}\rho E_{i_{1}...i_{N}}^{\dag},
\end{equation}
where $E_{i_{1}...i_{N}}$'s are Kraus operators which are defined as 
\begin{equation}
 E_{i_{1}...i_{N}}=\sqrt{P_{i_{1}...i_{N}}}A_{i_{1}} \otimes ... \otimes A_{i_{N}} , \quad \sum_{i}P_{i_{1}...i_{N}}=1.
\end{equation}
Here $P_{i_{1}...i_{N}}$ is the probability that a random sequence of operations is acted on the sequence input  $N$ qubits which are transmitted through the channel. In general, the Kraus operators for two consecutive uses of a two-qubit quantum channel can be represented as
\begin{equation}
E_{ij}=\sqrt{P_{ij}} A_{i}\otimes A_j.
\end{equation}
For uncorrelated channel we have $P_{ij} = P_iP_j$ and  Kraus
operators are independent of each other.  Whereas, for correlated channel based on Bayes rule we have $P_{ij}=P_iP_{j \vert i}$, where $P_{j \vert i}$ is the conditional probability. Thus, for two  consecutive uses of a two-qubit quantum channel with partial correlation the Kraus operators can be represented as 
\begin{equation}
E_{ij}=\sqrt{P_{i}[(1-\mu)P_{j}+\mu \delta_{ij}]},
\end{equation}
where $\mu \in [0,1]$ defines the correlation of the quantum channel.
According to the Kraus operator formalism the final state is given by
\begin{eqnarray}\label{dynamicsfinal}
\varepsilon(\rho)&=&(1-\mu)\sum_{i,j} E_{ij}\rho E_{ij}^{\dag}+ \mu \sum_{k}E_{kk}\rho E_{kk}^{\dag} \nonumber \\
&=&(1-\mu)\varepsilon_{un}(\rho)+\mu \varepsilon_{co}(\rho),
\end{eqnarray}
where $\varepsilon_{un}$ represents the uncorrelated channel and $\varepsilon_{co}$ stands for the correlated channel. In the case $\mu=0$, there is no correlation between two consecutive uses of channel and when $\mu=1$, the channel is fully correlated. In other words, $\mu=0$ represents the channel without memory and $\mu=1$ implies the channel with memory. Note that, in all parts of this work, we will consider the following initial states
\begin{equation}
\rho_0=r \vert \psi \rangle\langle \psi \vert + \frac{1-r}{4}\mathrm{I},
\end{equation}
where $\vert \psi \rangle = \sqrt{1-\alpha^{2}}\vert 01 \rangle + \alpha \vert 10 \rangle$, $0 \leq \alpha \leq 1$ and $r$ represents the purity of the initial state.

        
   
\subsection{Unital correlated  channel }
The completely positive trace preserving channel $\varepsilon$ is unital if it maps the identity operator $\sigma_{0}=\mathcal{I}$ to itself in the same space, i.e. $\varepsilon(\sigma_{0})=\sigma_{0}$. For single-qubit systems, the unital channel can be represented in terms of a convex combination of the Pauli operators \cite{Nielsen,Imre,King}. From geometrical insights, the unital channels map the center of the Bloch sphere to itself. Here we consider dephasing colored noise in the category of Pauli channels as an example of an unital channel \cite{Daffer}. We will study the QSL time  for correlated dephasing colored noise. \\
\textbf{Pure dephasing colored noise}: Let us consider the interaction between a single-qubit system and an environment which has the property of  random telegraph signal noise \cite{Daffer}. The dynamics of a single-qubit is described by the time-dependent Hamiltonian 
\begin{equation}
H(t)=\sum_{k=1}^{3} \Gamma_k(t) \sigma_k,
\end{equation}
where $\sigma_k$'s are the Pauli  operators in ($x,y,z$)directions, and $\Gamma_k(t)$'s are random variables
which follow the statistics of a random telegraph signal. $\Gamma_{k}(t)$ depends on the random variable  $n_{k}(t)$ as $\Gamma_{k} (t)=\alpha_k n_k(t)$, where $n_{k}(t)$ has a Poisson distribution with  an average value equal to $t/2\tau_k$ and $\alpha_k$'s are coin-flip random variables that can have values $\pm \alpha_k$ randomly.  We  have a dephasing model with colored noise if $\alpha_1=\alpha_2= 0$ and $\alpha_3=\alpha$. In this case, the dynamics of single-qubit system can be described by the following Kraus operators
\begin{equation}
E_{0} = \sqrt{P_0}\sigma_{0} , \quad E_{3}  = \sqrt{P_3}\sigma_{3}, \\
\end{equation}
with $P_0=1-z_t$ and $P_3=z_t$, where $z_t=\frac{1-\Lambda(t)}{2}$ and $\Lambda(t)=e^{-t/2\nu}[\cos(\mu t/2\nu)+\sin(\mu t/2\nu)/\mu]$, $\mu=\sqrt{(4 \nu )^{2}-1}$ . Here, the range of $\nu$ quantifies  the interval in which the channel is non-Markovian. Based on the results presented in Ref. \cite{Haseli1}, the quantum evolution will be non-Markovian  if $\nu \geq 1/4$.
\begin{figure}[!]  
\centerline{\includegraphics[scale=0.7]{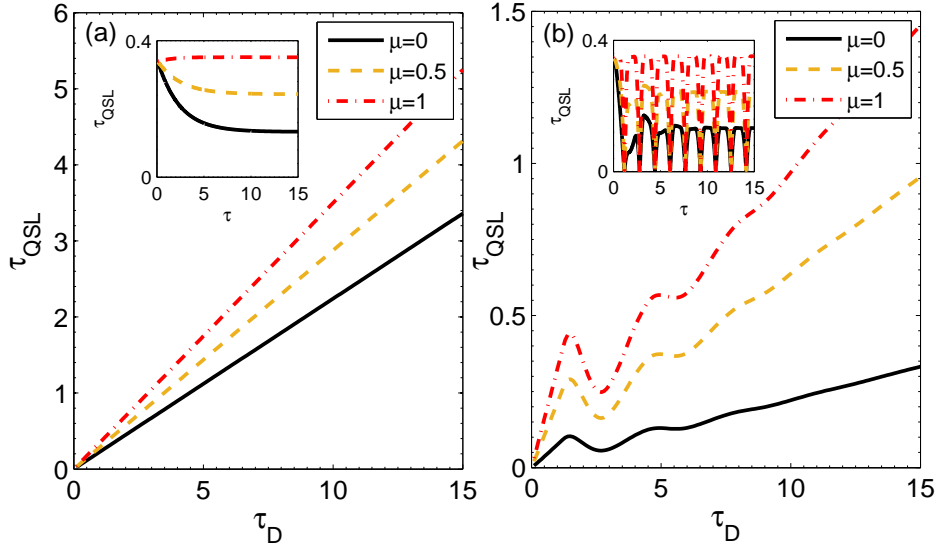}}
\caption{QSL time  for correlated pure dephasing colored as a function of the driving time $\tau_D$ when the initial state parameters are $r=1/2$ and $\alpha=1/\sqrt{2}$ and $\tau=1$. (a)The dynamics is Markovian $\nu=0.1$ (b)The dynamics is non-Markovian $\nu=1$. The inset represents the QSL time as a function of initial time $\tau$ when $\tau_D=1$. }\label{Fig1}
\end{figure}
When two-qubit are transmitted through the colored pure dephasing channel, the channel with uncorrelated noise can  be defined by the following Kraus operators
\begin{equation}
E_{ij}=\sqrt{P_i P_j}\sigma_{i} \otimes \sigma_{j}, \quad i,j \in \lbrace 0,3 \rbrace.
\end{equation}
In the presence of  correlation between  two successive uses of the colored pure dephasing channel on two-qubit system, the Kraus operators $E_{kk}$ are given by
\begin{equation}
E_{kk}=\sqrt{P_{k}}\sigma_{k}\otimes\sigma_{k}, \quad k \in \lbrace 0,3 \rbrace.
\end{equation}
 \textbf{From Eq. (\ref{dynamicsfinal}), the  elements of the time-dependent density matrix of a two qubit system under correlated  dephasing colored noise can be written as
\begin{eqnarray}
\rho^{t}_{11}&=&\rho^{t}_{44}=\frac{1-r}{4},  \nonumber \\
\rho^{t}_{22}&=&\frac{1}{4} \left(\left(3-4 \alpha ^2\right) 1+r\right), \nonumber \\
\rho^{t}_{33}&=&\frac{1}{4} \left(1-\left(1-4 \alpha ^2\right) r\right), \nonumber \\
\rho^{t}_{23}&=&\rho^{t\star}_{32}=\alpha  \sqrt{1-\alpha ^2} r \left(\mu +(1-\mu ) (1-z_t)^2\right). 
\end{eqnarray}
So, QSL time in Eq. (\ref{(QSL)T}) is derived as 
\begin{equation}
\tau_{QSL}=\frac{2 \alpha ^2\vert \left(1-\alpha ^2\right)  r^2 \left(\mu +( 1-\mu) (z_{\tau} -1)^2\right) \left(\left(z_{\tau+\tau_D} -1\right){}^2-(z_{\tau}-1)^2\right)\vert}{\frac{2 \sqrt{2} \alpha  \sqrt{1-\alpha ^2}  r}{\tau_D} \int_{\tau}^{\tau+\tau_D}(1-z_t) \dot{z}_t dt}
\end{equation}}
\textbf{In Fig. \ref{Fig1}, the QSL time is plotted as a function of driving time $\tau_D$ for correlated colored pure dephasing channel for different values of the correlation parameter $\mu$ when $\tau=1$. The insets represent the QSL time in terms of the initial time $\tau$ for different values of correlation parameter $\mu$ when $\tau_D=1$. In Fig. \ref{Fig1}(a) the environmental parameter is chosen such that the evolution is Markovian ($\nu=0.1$). As can be seen from Fig. \ref{Fig1}(a), the QSL time is increased by increasing the correlation parameter $\mu$. In Fig. \ref{Fig1}(b) the evolution is non-Markovian ($\nu=1$). As can be seen from Fig. \ref{Fig1}(b), the QSL time is also increased by increasing correlation parameter. From Figs. \ref{Fig1}(a) and \ref{Fig1}(b) one can find that for both Markovian and non-Markovian evolution the QSL time for correlated channel ($\mu=1$) is greater than uncorrelated channel ($\mu=0$). In other word, in the presence of correlation between two successive uses of the colored pure dephasing channel on two-qubit system the quantum evolution will be slower than the case in which the correlation does not exist.  }
\subsection{Non-unital correlated channel}
In this section we will study the QSL time for  non-unital correlated channels. Here, we consider correlated amplitude damping and correlated squeezed generalized amplitude damping channels as the examples for non-unital correlated channels.\\
\textbf{Correlated amplitude damping channel:}\\
Let us consider a two-level quantum system  that interacts with a zero temperature environment which is described by a collection of bosonic oscillators . In this model the  corresponding interaction Hamiltonian is given by
\begin{equation}
H=\omega_0 \sigma_+ \sigma_- + \sum_{k}\omega_{k}a_{k}^{\dagger}a_k +(\sigma_+ B + \sigma_- B^{\dagger}),
\end{equation}
where $\sigma_\pm$ are the raising and lowering operators of the two-level quantum system  having the transition frequency $\omega_0$ and $B=\sum_{k}g_{k}a_{k}$. $a_{k}$ and $a^{\dagger}_{k}$ are the annihilation and creation operators of the environment with
the frequencies $\omega_k$, respectively. Let us assume that the environment  has an spectral density of the form $J(\omega)=\gamma_{0}\lambda^{2}/2\pi\left[ (\omega_{0}-\omega)^{2}+\lambda^{2} \right] $, where the the coupling spectral width  $\lambda$ is connected to the correlation time of the environment $\tau_{B}$ by $\tau_B \sim 1/\lambda$. $\gamma_{0}$ is related to the relaxation time of the system  $\tau_{R}$ by $\tau_{R}\sim 1/\gamma_0$. The dynamics of the two-level quantum system, with this spectral density, can be described by  a master equation having the form of
\begin{equation}
\dot{\rho}_{t}=L_t\rho_t=\gamma_{t}\left(\sigma_- \rho_t \sigma_+ - \frac{1}{2} \left\lbrace \sigma_+\sigma_-,\rho_{t} \right\rbrace \right) ,
\end{equation}
where time-dependent decay rate is given by 
\begin{equation}
\gamma_t=\frac{2 \gamma_{0} \lambda \sinh (dt/2)}{d \cosh(dt/2)+\lambda \sinh(dt/2)}, \quad d=\sqrt{\lambda^2-2\gamma_0\lambda}.
\end{equation}
The dynamics of such a two-level quantum system can be expressed by the following Kraus operators as
\begin{equation}
A_{0}=\left(
\begin{array}{cc}
 \sqrt{1-p_t}  & 0 \\
 0 & 1 \\
\end{array}
\right), \quad A_1=\left(
\begin{array}{cc}
0  & 0 \\
\sqrt{p_t} & 0 \\
\end{array}
\right),
\end{equation}
where the parameter $p_t$ is given by
\begin{equation}
p_t=1-e^{- \lambda t}\left[ \cosh(\frac{d t }{2})+\frac{\lambda}{d}\sinh(\frac{d t }{2}) \right]^{2}.
\end{equation}

So, the quantum amplitude damping channel with uncorrelated noise can  be defined as 
\begin{equation}
E_ij=A_i \otimes A_j, \quad (i,j=0,1).
\end{equation}
The Kraus operators for two consecutive uses of a two-qubit  amplitude damping correlated quantum channel can be represented as
\begin{equation}
A_{00}=\left(
\begin{array}{cccc}
 \sqrt{1-p_t}  & 0 & 0 & 0 \\
 0 & 1 & 0 & 0 \\
 0 & 0 & 1 & 0 \\
 0 & 0 & 0 & 1 \\
\end{array}
\right), \quad A_{11}=\left(
\begin{array}{cccc}
 0  & 0 & 0 & 0 \\
 0 & 0 & 0 & 0 \\
 0 & 0 & 0 & 0 \\
 \sqrt{p_{t}} & 0 & 0 & 0 \\
\end{array}
\right)
\end{equation}
From Eq. (\ref{dynamicsfinal}), the evolving 
density matrix elements of a two qubit system under correlated  amplitude damping channel can be written as
\begin{eqnarray}
\rho^{t}_{11}&=&\frac{1}{4} (1-r) (1-p_t) (1-(1-\mu ) p_t), \nonumber \\
\rho^{t}_{22}&=&\frac{1}{4} \left(-4 \left(1-\alpha ^2\right) (1-\mu ) r p_t-(1-\mu ) (1-r) p_t p_t+\left(3-4 \alpha ^2\right) r+1\right), \nonumber \\
\rho^{t}_{33}&=&\frac{1}{4} \left(-4 \alpha ^2 (1-\mu ) r p_t-(1-\mu ) (1-r) p_t p_t-\left(1-4 \alpha ^2\right) r+1\right), \nonumber \\
\rho^{t}_{44}&=&\frac{1}{4} ((2-3 \mu ) r p_t+(1-\mu ) (1-r) p_t p_t-\mu  p_t+2 p_t-r+1), \nonumber \\
\rho^{t}_{23}&=&\rho^{t\star}_{32}=\alpha  \sqrt{1-\alpha ^2} r (1-(1-\mu ) p_t)
\end{eqnarray}
One can obtain the singular value of $\dot{\rho}_t$ as
\begin{eqnarray}
\Lambda_1&=&\frac{1}{2} (\mu -1) (r-1) p(t) \dot{p}_t, \nonumber \\
\Lambda_2 &=& \frac{1}{2} (\mu -1) (r p(t)-p(t)-2 r) \dot{p}_t, \nonumber \\
\Lambda_3&=&\frac{1}{4}  (-\mu +2 \mu  r p(t)-2 r p(t)-2 \mu  p(t)+2 p(t)-3 \mu  r+2 r+2)\dot{p}_t, \nonumber \\
\Lambda_4&=& \frac{1}{4}  (\mu +2 \mu  r p(t)-2 r p(t)-2 \mu  p(t)+2 p(t)-\mu  r+2 r-2)\dot{p}_t.
\end{eqnarray}
Now from Eq. \ref{(QSL)T}, one can obtain the QSL time for correlated amplitude damping quantum channel.
\begin{figure}[!]  
\centerline{\includegraphics[scale=0.7]{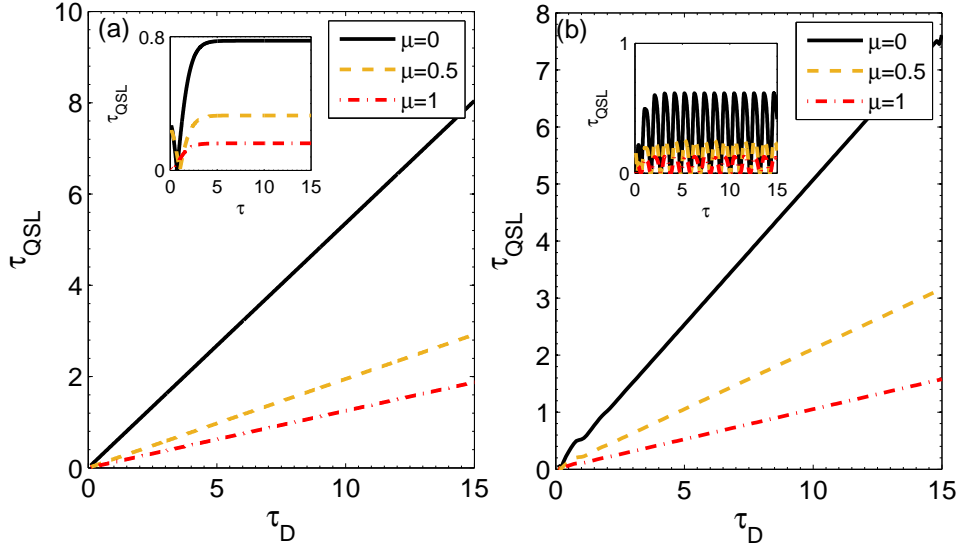}}
\caption{QSL time  for correlated amplitude damping channel as a function of the driving time $\tau_D$ when the initial state parameters are $r=1/2$ and $\alpha=1/\sqrt{2}$ and $\tau=1$. (a)The dynamics is Markovian $\lambda/\gamma_{0}=2$(b)The dynamics is non-Markovian  $\lambda/\gamma_0=0.2$. The inset represents the QSL time as a function of initial time $\tau$ when $\tau_D=1$. }\label{Fig2}
\end{figure}

\textbf{In Fig. \ref{Fig2}, the QSL time is plotted as a function of driving time $\tau_D$ for correlated amplitude damping channel with different values of correlation parameter $\mu$ when $\tau=1$. The inset shows the quantum speed limit time in terms of the initial time $\tau$ for different values of correlation parameter $\mu$ when $\tau_D=1$. In Fig. \ref{Fig2}(a) we choose $\lambda / \gamma_0 =2$ and the evolution is Markovian.  As can be seen from Fig. \ref{Fig2}(a), the QSL time is decreased by increasing the correlation parameter $\mu$. In Fig. \ref{Fig2}(b) we have $\lambda /\gamma_0 =0.2$ and the evolution is non-Markovian. As can be seen from Fig. \ref{Fig2}(b), the QSL time is also decreased by increasing correlation parameter. From Figs. \ref{Fig1}(a) and \ref{Fig2}(b) one can find that for both Markovian and non-Markovian evolution the QSL time for correlated channel ($\mu=1$) is smaller than uncorrelated channel ($\mu=0$). In other word, in the presence of correlation between two successive uses of the colored pure dephasing channel on two-qubits system the quantum evolution will be faster than the case in which the correlation does not exist.  }\\
\textbf{Correlated Squeezed Generalized Amplitude Damping Channels}:  \\
An amplitude damping channel represents a physical process such as energy dissipation of a two-level quantum system due to spontaneous emission of a photon into the vacuum at zero temperature \cite{Nielsen}. The generalized amplitude damping (GAD) channel  describes the relaxation of a quantum system when the surrounding environment is at finite temperature  initially i.e., when the environment starts from a mixed state \cite{Fujiwara}. Generalized amplitude damping channel  is developed as a squeezed generalized amplitude damping (SGAD) channel by considering a squeezed thermal environment \cite{Srikanth}. The SGAD channel  is a combination of both effects of dissipation at finite temperature and environment squeezing \cite{Daffer1,Banerjee,Wilson,Banerjee1}.

The SGAD channel defines the quantum noise in which the quantum system interacts with an environment that is initially in a squeezed thermal state under the Markov and Born approximations. The dynamics of such a quantum system can be described by following Lindblad master equation
\begin{eqnarray}\label{master}
L(\rho_{t})&=&-\frac{\Omega (n+1)}{2}(\sigma_{+}\sigma_{-}\rho_{t} + \rho_{t} \sigma_{+}\sigma_{-} - 2 \sigma_{-} \rho_{t} \sigma_{+})\nonumber \\
&-&\frac{\Omega n}{2}(\sigma_{-}\sigma_{+}\rho_{t} + \rho_{t} \sigma_{-}\sigma_{+} - 2 \sigma_{+} \rho_{t} \sigma_{-}) \nonumber \\
&-&\Omega m (\sigma_{+}\rho_t \sigma_{+} + \sigma_{-}\rho_t \sigma_{-}),
\end{eqnarray}
where $\sigma_{+}=\frac{1}{2}(\sigma_{1}+i \sigma_{2})$ and $\sigma_{-}=\frac{1}{2}(\sigma_{1}-i \sigma_{2})$ are creation and annihilation operators, respectively, $n$ is associated with the number of thermal photons, $m$ is the squeezing parameter ($m$ and $n$ satisfy $m<n+1/2$) and $\Omega$ is the  dissipation rate related to  spontaneous emission at zero-temperature \cite{Breuer,Nielsen,Daffer1}. If $m=0$ then SGAD transforms to the GAD channel. When $m=n=0$ the SGAD channel reduces to the amplitude damping channel.
\begin{figure}[!]
\centerline{\includegraphics[scale=0.7]{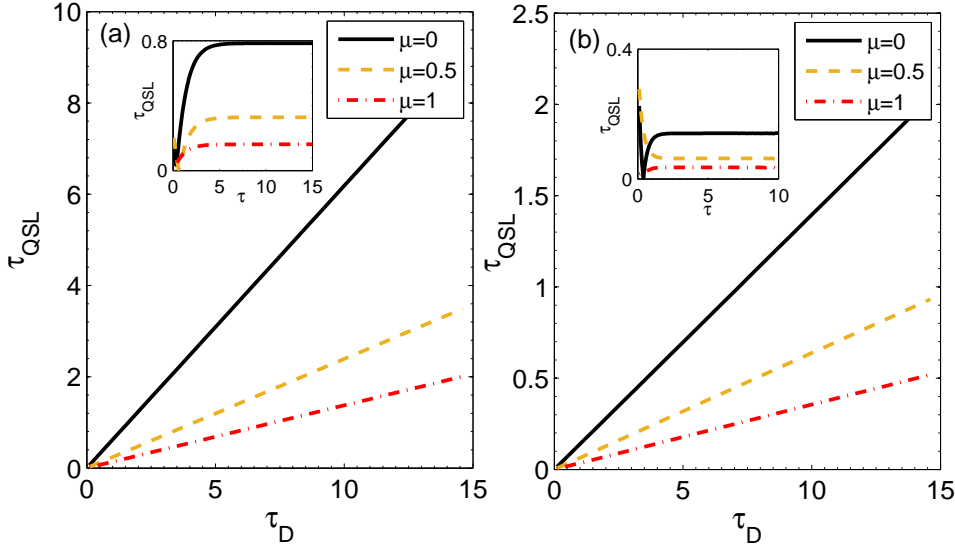}}
\caption{QSL time as a function of the driving time $\tau_D$ when the initial state parameters are $r=1/2$ and $\alpha=1/\sqrt{2}$ when $\tau=1$ for (a) Generalized amplitude damping channel with $m=0,n=1$ and (b) squeezed Generalized amplitude damping channel with $m=1,n=1$. The inset represents the QSL time as a function of initial time $\tau$ when $\tau_D=1$. }\label{Fig3}
\end{figure}
We first review the simple method for  solving the master equation in Eq. (\ref{master}) to find the structure of uncorrelated SGAD channel. The dynamics of single-qubit state with initial input $\rho_{0}=\sum_{i,j=0}^{1}\vert i \rangle \langle j \vert$, associated with this Lindblad master equation has the following form \cite{Daffer1}
\begin{eqnarray}\label{30}
\rho_{t} &=& e^{L t} \rho \nonumber \\ 
&=&\sum_{i}tr(\mathcal{R}_i \rho)e^{\eta_{i}t}\mathcal{L}_{i}=\sum_{i}tr(\mathcal{L}_i \rho)e^{\eta_{i}t}\mathcal{R}_{i},
\end{eqnarray} 
where $\mathcal{R}_i$ and $\mathcal{L}_i$ are the  right and left eigenoperators of super-operator $L$ in Lindblad master equation, $\eta_{i}$'s are corresponding eigenvalues of these eigenoperators, such that
\begin{equation}\label{31}
L \mathcal{R}_{i}=\eta_{i}\mathcal{R}, \quad  \mathcal{L}_{i}L=\eta_{i}\mathcal{L}_{i},
\end{equation}
with $tr(\mathcal{L}_{i}\mathcal{R}_j)=\delta_{ij}$. For SGAD channel, $\mathcal{R}_{i}$'s and  $\mathcal{L}_{i}$'s can be defined as \cite{Jeong}
\begin{eqnarray}\label{32}
\mathcal{R}_{1}&=&\frac{1}{\sqrt{2}}(\mathbf{I}_{2 \times 2} - \frac{1}{2n+1}\sigma_3), \quad \mathcal{L}_{1}=\frac{1}{\sqrt{2}}\mathbf{I}_{2 \times 2}, \nonumber \\
\mathcal{R}_{2}&=&\mathcal{L}_{2}=\frac{1}{\sqrt{2}}(\sigma_+ + \sigma_-),\nonumber \\
\mathcal{R}_{3}&=&-\mathcal{L}_{3}=\frac{1}{\sqrt{2}}(\sigma_- - \sigma_+), \nonumber \\
\mathcal{R}_{4}&=& \frac{1}{\sqrt{2}} \sigma_{3}, \quad  \mathcal{L}_{4}= \frac{1}{\sqrt{2}}(\frac{1}{2n+1} \mathbf{I}_{2 \times 2} + \sigma_3).
\end{eqnarray}
The eigenvalues $\eta_{i}$ are given by
\begin{eqnarray}\label{33}
\eta_{1}&=&0, \quad \eta_{2}=-\Omega(n+m+\frac{1}{2}), \nonumber \\
\eta_{3}&=&-\Omega(n-m+\frac{1}{2}), \quad \eta_{4}=-2\Omega(n+\frac{1}{2}).
\end{eqnarray}
From Eqs. (\ref{30}), (\ref{31}), (\ref{32}) and (\ref{33}), the evolved single-qubit density matrix can be quantified as 
\begin{equation}\label{dynamics}
\rho_{t}=
  \left(   {\begin{array}{cc}
  \frac{n+p_t^{2}(n+1)\rho_{11}-n \rho_{22}}{2n+1} & p_t(q_{t}\rho_{12}-r_{t}\rho_{21}) \\
  p_t(q_{t}\rho_{21}-r_{t}\rho_{12}) & 1- \frac{n+p_t^{2}(n+1)\rho_{11}-n \rho_{22}}{2n+1} \\
  \end{array} } \right),
\end{equation}
where $p_t=e^{-\Omega(n+1/2)t}$, $q_t=\cosh(\Omega m t)$ and $r_t=\sinh(\Omega m t)$. From Eq. (\ref{dynamics}), the Kraus operators $A_i$ for single-qubit dynamics under SGAD channel can be written as
\begin{eqnarray}\label{kraussingle}
A_{1}&=&
  \left(   {\begin{array}{cc}
   \sqrt{\frac{n}{2n+1}+\frac{n+1}{2n+1} p_{t}^{2}- p_{t}q_{t}} & 0 \\
  0 & 0 \\
  \end{array} } \right), \nonumber \\
  A_{2}&=&
  \left(   {\begin{array}{cc}
   0 & 0 \\
  \sqrt{\frac{n+1}{2n+1}(1-p_{t}^2)-p_{t}r_{t}} & 0 \\
  \end{array} } \right), \nonumber \\
   A_{3}&=&
  \left(   {\begin{array}{cc}
   0 & 0 \\
  0 & \sqrt{\frac{n+1}{2n+1}+\frac{n}{2n+1}p_{t}^2-p_{t}r_{t}} \\
  \end{array} } \right), \nonumber \\
    A_{4}&=&
  \left(   {\begin{array}{cc}
   \sqrt{p_{t} q_{t}} & 0 \\
  0 & \sqrt{p_{t}q_{t}} \\  \end{array} }\right), \nonumber \\
  A_{5}&=&
  \left(   {\begin{array}{cc}
  0 &  \sqrt{p_{t}r_{t}} \\
   \sqrt{p_{t}r_{t}}  & 0\\  \end{array} }\right), \nonumber \\
    A_{6}&=&
  \left(   {\begin{array}{cc}
  0 &  \sqrt{\frac{n}{2n+1}(1-p_{t}^{2})-p_t r_t} \\
  0  & 0\\  \end{array} }\right).
\end{eqnarray}
Now, we consider two consecutive uses of the SGAD channel on two-qubit quantum system. Uncorrelated SGAD channel $\varepsilon_{un}$ can  be shown by the following Kraus operators \cite{Jeong}
\begin{equation}
E_{ij}= A_{i} \otimes A_{j}, \quad i,j=1,...,6,
\end{equation}
We consider the following correlated Lindblad master equation for two-qubit system to find the structure of the correlated SGAD channel   
\begin{eqnarray}\label{master2}
\tilde{L}(\rho_{t})&=&-\frac{\Omega (n+1)}{2}(\sigma_{+}^{\otimes 2}\sigma_{-}^{\otimes 2}\rho_{t} + \rho_{t} \sigma_{+}^{\otimes 2}\sigma_{-}^{\otimes 2} - 2 \sigma_{-}^{\otimes 2} \rho_{t} \sigma_{+}^{\otimes 2})\nonumber \\
&-&\frac{\Omega n}{2}(\sigma_{-}^{\otimes 2}\sigma_{+}^{\otimes 2}\rho_{t} + \rho_{t} \sigma_{-}^{\otimes 2}\sigma_{+}^{\otimes 2} - 2 \sigma_{+}^{\otimes 2} \rho_{t} \sigma_{-}^{\otimes 2}) \nonumber \\
&-&\Omega m (\sigma_{+}^{\otimes 2}\rho_t \sigma_{+}^{\otimes 2} + \sigma_{-}^{\otimes 2}\rho_t \sigma_{-}^{\otimes 2}),
\end{eqnarray}
where $\sigma_{\pm}^{\otimes 2}=\sigma_{\pm} \otimes \sigma_{\pm}$. The correlated dynamics of a two-qubit state can be found to be similar to the single-qubit case by using Eqs.(\ref{30}) and (\ref{31}). We consider a general two-qubit state $\rho_0=\sum_{\alpha_1,\alpha_2=1}^{4} \rho_{\alpha_1,\alpha_2}\vert \alpha_{1} \rangle \langle \alpha_{2} \vert$ as an initial input state, where $\vert \alpha_{1,2} \rangle \in \left\lbrace \vert 00 \rangle, \vert 01 \rangle, \vert 10 \rangle, \vert 11 \rangle \right\rbrace $. The solution of correlated master equation in Eq.\ref{master2} is derived as 
\begin{eqnarray}\label{38}
\rho_{11}(t)&=&\frac{1}{2n+1} \left( \left((n+1) p_t^2-(2 n+1) s_t \left(1-u_t\right)+n\right) \rho _{11} \right. \nonumber \\
&+& \left.  \left(n-p_t \left(n p_t+2 (2 n+1) r_t\right)\right) \rho _{44} \right), \nonumber \\
\rho_{12}(t)&=& \sqrt{s_t u_t} \rho _{12} ,   \nonumber \\ 
\rho_{13}(t)&=& \sqrt{s_t u_t} \rho _{13} , \nonumber \\
\rho_{14}(t)&=& (\sqrt{s_t} u_t - p_t \left(1-q_t\right)) \rho _{14} -p_t  r_t \rho _{41}, \nonumber \\
\rho_{22}(t)&=& \rho_{22}, \nonumber \\
\rho_{23}(t)&=&\rho_{23}, \nonumber \\
\rho_{24}(t)&=& \sqrt{u_t}\rho _{24}, \nonumber \\
\rho_{33}(t)&=&\rho _{33}, \nonumber \\
\rho_{34}(t)&=& \sqrt{u_t}\rho _{34}, \nonumber \\
\rho_{44}(t)&=&1-\rho_{11}(t)-\rho_{22}-\rho_{33}.
\end{eqnarray}
\vfill
From Eq. (\ref{38}), the Kraus operators $E_{kk}$ for correlated part is obtained as
 \begin{eqnarray}\label{kraustwo}
  E_{11}&=&
  \left(   {\begin{array}{cccc}
  \sqrt{s_{t}} & 0 & 0 & 0\\
  0 & 1 & 0 & 0\\
  0 & 0 & 1 & 0\\
  0 & 0 & 0 & \sqrt{u_t}\\
  \end{array} } \right), \nonumber \\
 E_{22}&=&
  \left(   {\begin{array}{cccc}
  0 & 0 & 0 & 0\\
  0 & 0 & 0 & 0\\
  0 & 0 & 0 & 0\\
  \sqrt{\frac{n+1}{2 n+1}(1-p_{t}^{2})-p_{t}r_{t}}  & 0 & 0 & 0\\
  \end{array} } \right), \nonumber \\  
   E_{33}&=&
  \left(   {\begin{array}{cccc}
  0 & 0 & 0 & \sqrt{\frac{n}{2n+1}(1-p_{t}^{2})-p_{t}r_{t}}\\
  0 & 0 & 0 & 0\\
  0 & 0 & 0 & 0\\
  0 & 0 & 0 & 0\\
  \end{array} } \right), \nonumber \\   
 E_{44}&=&
  \left(   {\begin{array}{cccc}
  \sqrt{\frac{n}{2n+1}+\frac{n+1}{2n+1}p_{t}^{2}-p_{t}(q_{t}-1)-s_{t}} & 0 & 0 & 0\\
  0 & 0 & 0 & 0\\
  0 & 0 & 0 & 0\\
  0 & 0 & 0 & 0\\
  \end{array} } \right), \nonumber \\
   E_{55}&=&
  \left(   {\begin{array}{cccc}
  0 & 0 & 0 & 0\\
  0 & 0 & 0 & 0\\
  0 & 0 & 0 & 0\\
  0 & 0 & 0 & \sqrt{\frac{n+1}{2n+1}+\frac{n}{2n+1}p_{t}^{2}-p_{t}(q_{t}-1)-u_t}\\
  \end{array} } \right), \nonumber \\ 
   E_{66}&=&
  \left(   {\begin{array}{cccc}
  \sqrt{p_{t}(q_t)-1} & 0 & 0 & 0\\
  0 & 0 & 0 & 0\\
  0 & 0 & 0 & 0\\
  0 & 0 & 0 & \sqrt{p_{t}(q_t)-1}\\
  \end{array} } \right), \nonumber \\ 
   E_{77}&=&
  \left(   {\begin{array}{cccc}
  0 & 0 & 0 & i \sqrt{p_{t}r_{t}}\\
  0 & 0 & 0 & 0\\
  0 & 0 & 0 & 0\\
  i \sqrt{p_{t}r_{t}} & 0 & 0 & 0\\
  \end{array} } \right), 
 \end{eqnarray}
where $u_{t}=e^{-\Omega n t}$ and $s_{t}=e^{-\Omega (n+1)t}$. From Eq. (\ref{(QSL)T}), one can find the QSL time for correlated SGAD quantum channel after some straightforward calculation with large output . \textbf{In Fig. \ref{Fig3}, the QSL time is plotted as a function of driving time $\tau_D$ for correlated  GAD channel and correlated SGAD channel with different values of correlation parameter $\mu$ when $\tau=1$. The inset shows the quantum speed limit time in terms of the initial time $\tau$ for different values of correlation parameter $\mu$ when $\tau_D=1$. In Fig. \ref{Fig3}(a) QSL time is plotted as a function of driving time for correlated generalized amplitude damping i.e. the correlated channel with parameters $m=0, n=1$. As can be seen from Fig. \ref{Fig3}(a), the QSL time is decreased by increasing the correlation parameter $\mu$. In Fig. \ref{Fig3}(b) QSL time is plotted as a function of driving time $\tau_D$ for correlated squeezed generalized amplitude damping i.e. the correlated channel with parameters $m=1, n=1$. As can be seen from Fig. \ref{Fig3}(b), the QSL time is also decreased by increasing correlation parameter. From Figs. \ref{Fig3}(a) and \ref{Fig3}(b) one can find that for both GAD and SGAD correlated channels  the QSL time is decreased by increasing correlation parameter $\mu$. In other word, in the presence of correlation between two successive uses of these channels on two-qubits system the quantum evolution will be faster than the case in which the correlation does not exist.  }\\
 \section{Summary and conclusion}\label{Sec4}
We have studied the QSL time for correlated quantum channels, where the term correlated indicates the existence of correlations between two consecutive uses of the quantum channel on a two-qubit quantum system.  We have used correlated pure dephasing colored noise as an example for the unital correlated quantum channels and correlated amplitude damping and SGAD  channels as the examples for the non-unital quantum channels.\textbf{ We found that in the case of correlated dephasing colored noise for both Markovian and non-Markovian evolution the QSL time is increased by increasing correlation parameter  of quantum channel $\mu$. In other word, in the presence of correlation between two successive uses of the colored pure dephasing channel on two-qubits quantum system the quantum evolution will be slower than the case in which the correlation does not exist. In the case of correlated amplitude damping channel for both Markovian and non-Markovian evolution the QSL time is decreased by increasing correlation parameter  of quantum channel $\mu$. In the case of correlated generalized amplitude damping and correlated squeezed generalized amplitude damping channel the quantum speed limit time is decreased by increasing correlation parameter of quantum channel.  }  
\section*{Acknowledgments}
The authors would like to thank Prof. Masashi Ban for his valuable comments.

\end{document}